\documentclass[conference]{IEEEtran}

\usepackage{multirow}
\usepackage{subfigure}

\usepackage[pdftex]{graphicx}
\usepackage{amsmath}
\usepackage{todonotes}

\ifCLASSINFOpdf
 
\else

\fi
 
\begin{document}

\title{Real-time Analog Pixel-to-pixel Dynamic Frame Differencing with Memristive Sensing Circuits}

\author{\IEEEauthorblockN{Olga Krestinskaya, and  Alex Pappachen James}
\IEEEauthorblockA{School of Electrical and Computer Engineering\\
Nazarbayev University, 
Astana, Kazakhstan\\
Email: \{ok,~apj\}@ieee.org}
}

\maketitle

\begin{abstract}
In this paper, we propose an analog pixel differencing circuit for differentiating pixels between frames directly from CMOS pixels. The analog information processing at sensor is a topic of growing appeal to develop edge AI devices. The proposed circuit is integrated into a pixel-parallel and pixel-column architectures. The proposed system is design using TSMC $180nm$ CMOS technology. The power dissipation of the proposed circuit is $96.64mW$, and on-chip ares is $531.66 \mu m^2$. The architectures are tested for moving object detection application. 
\end{abstract}

\IEEEpeerreviewmaketitle

\section{Introduction}

 
  In the conventional real time object detection and tracking systems, the analog signals from the sensor is converted to digital domain using analog-to-digital converter (ADC) before the pixel differences are calculated. The sampling rates of the ADC, the need to store the pixels in a memory, and the circuit delays from the pixel arrays reduces the processing speed. The co-processing systems are digital and very often built on FPGA \cite{4600908,8070194,6949967} to have near sensor computing. An alternative to this approach is to move the processing to sensor in analog domain before the ADC stage. The continuous domain processing of pixel information potentially allows to improve the speed, and capture motions of ultra high speed objects.  However, the high speed processing of the signals from pixel sensors in analog domain is an open problem. This problem can be addressed using memristor based analog circuits.
Memristor has been proven to be useful for various object detection systems  \cite{7562381,access}, edge detection \cite{7544265} and image filtering applications \cite{irmanova2017neuromorphic,james2017unified}.

In this paper, we propose programmable analog near CMOS pixel sensor processing unit useful for various visual data processing applications.  The proposed circuit can be used for various similarity calculations tasks in static and dynamic systems. The memristor is used to control between the static and dynamic application of the circuit.
The static application include filtering operations performed on a single image. The dynamic application include the processing and comparison of several images in time useful for high speed motion detection and edge detection. In particular, this paper demonstrates high speed object detection application.


\section{Pixel Frame Differencing Circuit}

\subsection{Circuit design}

\begin{figure}
\centering
\includegraphics[width=50mm]{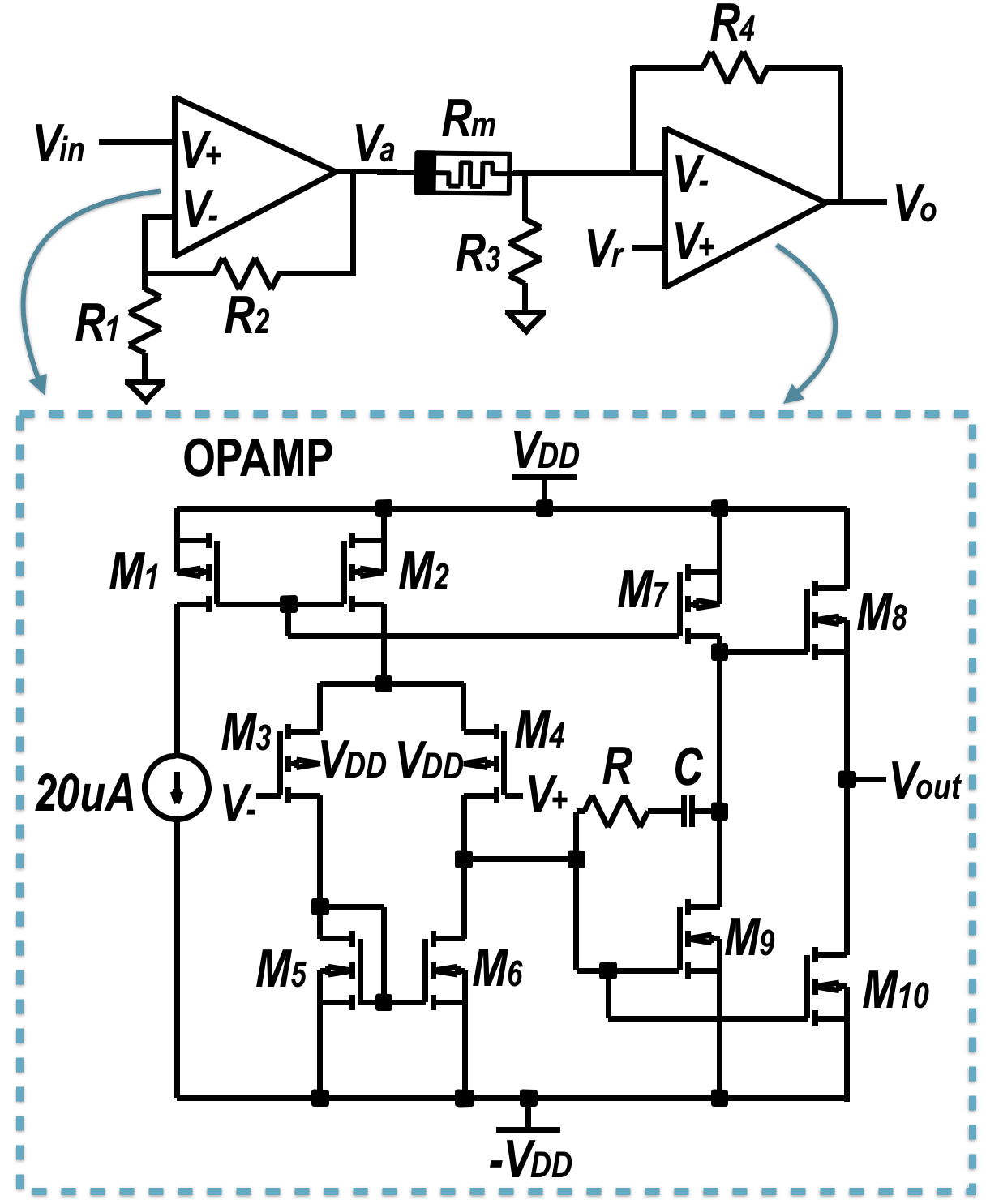}
\caption{Proposed circuit design.}
\label{f1}
\end{figure}

Fig. \ref{f1} illustrate the proposed circuit. Eq. \ref{eq1} show the output voltage of the circuit.
where $V_{in}$ and $V_r$ are the input pixels and $V_a=V_{in}(1+R_2/R_1)$. 

\begin{equation}
\label{eq1}
V_o=V_r(\frac{R_4}{R_m}+\frac{R_4}{R_3}+1)-V_a\frac{R_4}{R_m}
\end{equation}

The memristor $R_m$ is used to control the functionality of the circuit. The memristor can be programmed to either low resistance state $R_{on}$ or high resistance state $R_{off}$. We set $R_4=R_3=R_1=R$, $R_2=2R_1=2R$, $R_{on}=R$ and $R_{off}=100R$.
In case of $R_m=R_{on}$, $V_o=3V_r-3V_{in}$, which corresponds to the calculation of the difference between two pixels. In case of $R_m=R_{off}$, $V_o=2.01V_r-0.03V_{in}$, which corresponds to the preservation of the $V_r$ pixel.

The system was designed using TSMC $180 \mu m$ CMOS technology. The operational amplifier (OpAmp) parameters are the following: $V_{DD}=4V$, $M_1=0.18\mu/3\mu$, $M_2=0.18\mu/30\mu$, $M_3=M_4=0.18\mu/36\mu$, $M_5=M_6=0.18\mu/6\mu$, $M_7=0.18\mu/45\mu$, $M_8=M_9=M_{10}=0.18\mu/18\mu$ and the capacitors $C=1pF$ and $R=1k\Omega$. The resistor values are $R_1=1k\Omega$, $R_2=2k\Omega$, $R_3=1k\Omega$ and $R_4=1k\Omega$. The high resistance state of memristor $R_m$ is $R_{off}=100k\Omega$, and low resistance state is $R_{on}=1k\Omega$.

\subsection{System design}

\subsubsection{Static system}

For the static systems, the proposed circuit is placed inside each CMOS pixel sensor \cite{7539106}, making it a pixel parallel processing. The memristor serves as a switch between the difference operation and for obtaining the pure background information, which makes the system programmable. The background information is supplied to $V_r$. If the memristor is programmed to high resistance $R_{off}$, we observe purely background information provided by $V_r$. If the memristor is programmed to a low resistance $R_{on}$, we obtain the difference between the background and input pixel $V_{in}$.

\subsubsection{Dynamic system}

\begin{figure}[t]
\centering
\includegraphics[width=60mm]{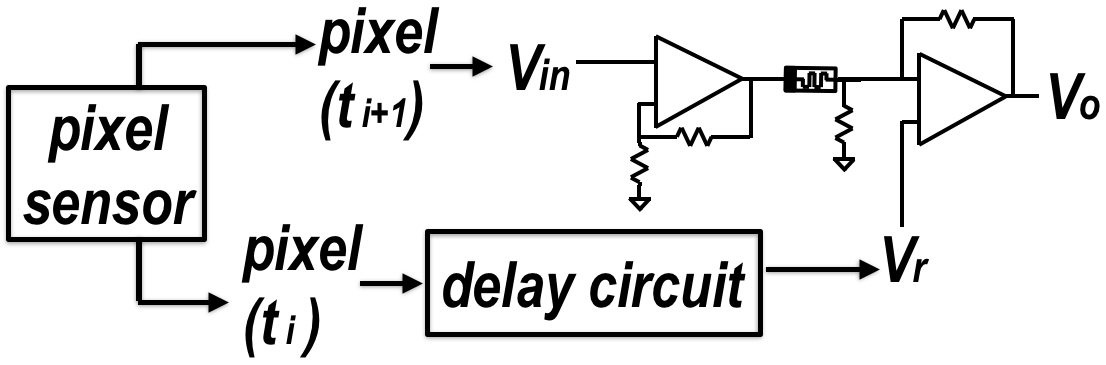}
\caption{System level design for dynamic system.}
\label{f2}
\end{figure}

In the dynamic system, the pixel difference in time is obtained. The dynamic system configuration is shown in Fig. \ref{f2}.
The inputs $V_r$ and $V_{in}$ are the inputs from the subsequent frames in time. The capacitive effect in the delay circuit provides a programmable delay to the system. As the system is analog, the output of the circuit produces a difference calculation in time without sampling the input and conversion to digital domain. In such system, each pixel should be connected to a separate proposed circuit to ensure parallel processing of the images and avoid implementation of an additional memory unit. 

\begin{figure}[t]
\centering
\includegraphics[width=90mm]{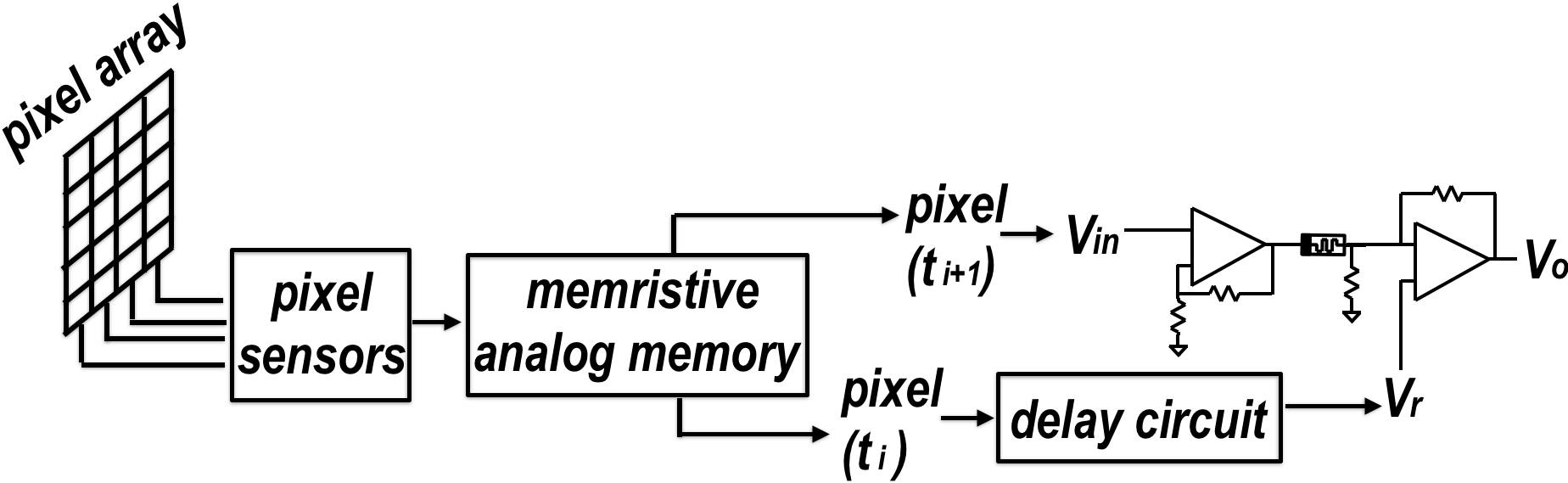}
\caption{System level design for dynamic system and reduced number of components.}
\label{f3}
\end{figure}

However, for the large images, the trade-off between processing speed and power consumption of the circuit should be achieved.  The system shown in Fig. 
\ref{f3} can reduce the power consumption of a pixel parallel approach. In this system, the rows of the image are read sequentially and stored in memristive analog memory unit. This approach allow to reduce the number of required circuit components. For the pixel array size of $n\times m$, where $n$ is a number of rows and $m$ is a number of columns in the image, the number of required components to implement the overall system can be reduced by $(1-1/m)\times 100\%$. 

\section{Simulation results}
\begin{figure*}[!t]
    \centering        
    \subfigure[]
    	{
    \includegraphics[width=180mm]{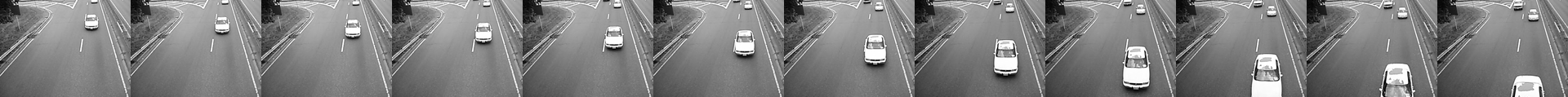}
		}       
     \subfigure[]
		{    	\includegraphics[width=180mm]{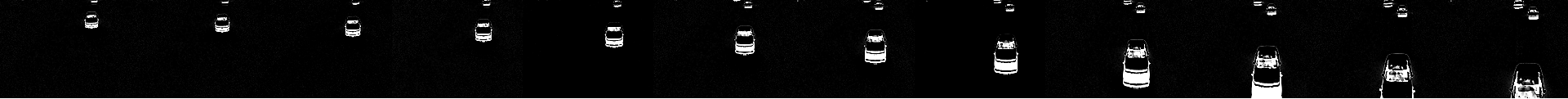}
		}
             \subfigure[]
		{    	\includegraphics[width=180mm]{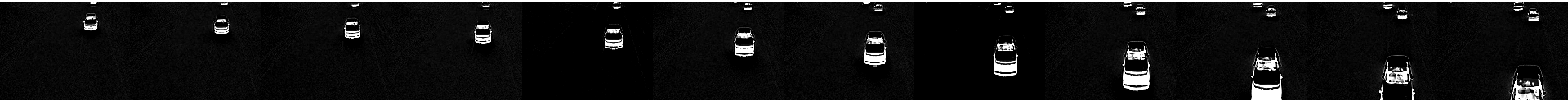}
		}
             \subfigure[]
		{    	\includegraphics[width=180mm]{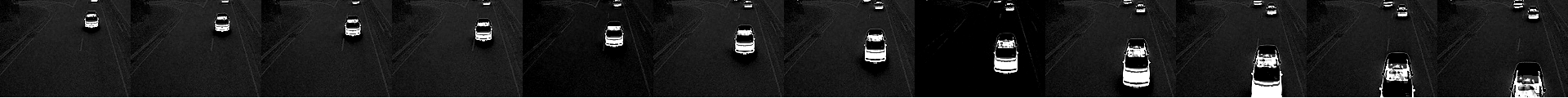}
		}
             \subfigure[]
		{    	\includegraphics[width=180mm]{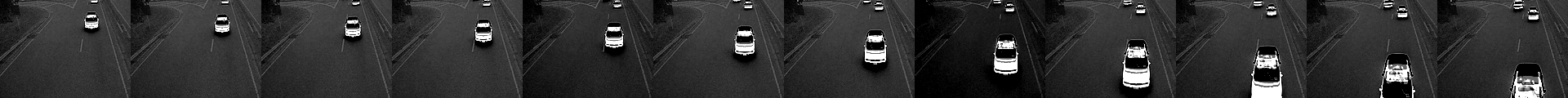}
		}
        \caption{System level simulations of the proposed method for objection: (a) original image and detected object for (b) ideally programmed memristor, (c) 10\% variation of memristor resistance, (d) 30\% variation in memristor resistance, and (e) 50\% variation of the resistance state of the memristor.}
        \label{f4}
\end{figure*}

The system level simulations for moving object detection have been performed using MATLAB scripts taking realistic SPICE device models for characterizing the underlying functional blocks. The simulated system corresponds to the case shown in Fig. \ref{f2}, where each image pixel contains a separate detector circuit.
The simulation results for the dynamic system are shown in Fig. \ref{f4}. Fig. \ref{f4} (a) corresponds to the actual sequence of images. Fig. \ref{f4} (b) illustrates the detection of the moving objects for the case of ideal memristor. However, in most of the cases, the memristor can have variations in $R_{on}$ and $R_{off}$ values. Therefore, we performed the simulations with the mismatch in the final memristor value. Fig. \ref{f4} (c), Fig. \ref{f4} (d) and Fig. \ref{f4} (e) show the object detection results for the random variation of resistance of the memristor for 10\%, 30\% and 50\%, respectively. The simulation results show that for the mismatch in the value of the memristor up to 30\%, the object is moving object is clear and it is possible to detect the moving object without additional filtering operations. For highly inaccurate memristors (approximately 50\% mismatch in the resistance values), the object can still be detected applying additional filtering operation. 

\begin{figure*}[t]
    \centering        
    \subfigure[]
    	{
    \includegraphics[width=45mm]{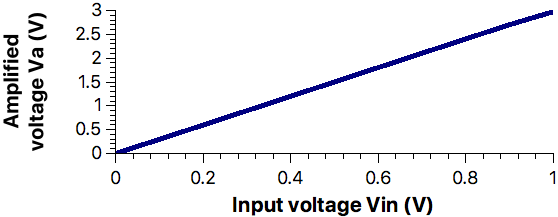}
		}       
     \subfigure[]
		{    	\includegraphics[width=45mm]{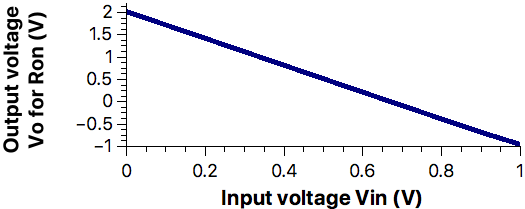}
		}
             \subfigure[]
		{    	\includegraphics[width=45mm]{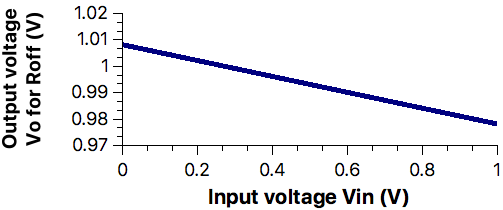}
		}
        \caption{Circuit simulations for difference input $V_{in}$ with $V_{r}=1V$: (a) $V_a$ voltage, (b) output voltage $V_o$ for $R_{on}$ memristor state, and  (c) $V_o$ for $R_{off}$ memristor state.}
        \label{f5}
\end{figure*}

\begin{figure*}[t]
    \centering        
    \subfigure[]
    	{
    \includegraphics[width=42mm]{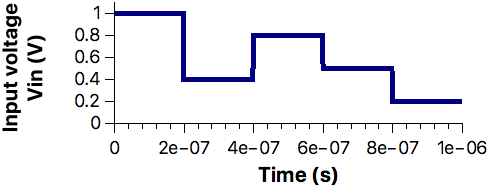}
		}       
     \subfigure[]
		{    	\includegraphics[width=42mm]{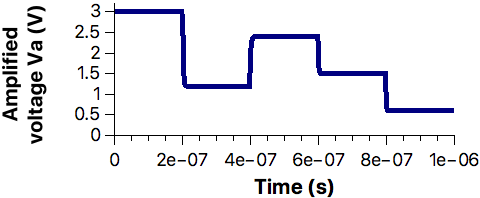}
		}
             \subfigure[]
		{    	\includegraphics[width=42mm]{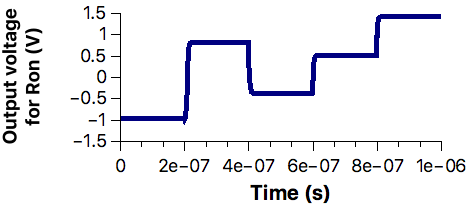}
		}
             \subfigure[]
		{    	\includegraphics[width=42mm]{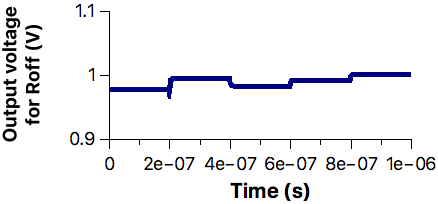}
		}
        \caption{Transient analysis of the circuit for $V_{r}=1V$: (a) input voltage $V_{in}$ voltage, (b) amplified input voltage $V_{a}$, (c) output voltage $V_o$ for $R_{on}$ memristor state, and (d) $V_o$ for $R_{off}$ memristor state.}
        \label{f6}
\end{figure*}

The circuit level simulations have been performed in SPICE and results are shown in Fig. \ref{f5}-\ref{f6}. Fig. \ref{f5} shown the amplitude of the signal for different values of $V_{in}$ and $V_r=1V$. Fig \ref{f5} (a) illustrates the amplified voltage $V_{a}$ after the first OpAmp (Fig. \ref{f1}). Fig. \ref{f5} (b) shows the output voltage $V_o$ for $R_m=R_{on}$, and Fig. \ref{f5} (c) illustrate the output voltage variation for $R_m=R_{off}$. There is a variation for the high resistance of the memristor, however variation is insignificant and can be neglected, as it will not effect the output background image.

The transient analysis is shown in Fig. \ref{f6}. Fig. \ref{f6} (a) illustrate an example of an input pixel signal in time from the object under study. Fig. \ref{f6} (b) corresponds to the amplified input $V_a$, and Fig. \ref{f6} (c) and Fig. \ref{f6} (d) shows the output signal $V_{o}$ corresponding to $R_m=R_{on}$ and $R_m=R_{off}$, respectively. The power dissipation of the proposed circuit is $96.64mW$, and on-chip ares is $531.66 \mu m^2$.

\section{Discussion}

The proposed analog domain signal processing near sensor system uses fewer number of circuit components in comparing to similar digital and mixed signal solutions for motion detection \cite{1,2,3}; the number of transistors and complexity of the circuits in the proposed analog design is significantly reduced.
One of the main application of the proposed system is the background subtraction in analog domain near to the sensor, reducing the need to have complex designs for ADC, digital memory array and control circuits.
The other application of the proposed system is an average similarity calculation and multiple pixel comparison for a static systems. In this process, the central pixel is compared to all nearby pixels and the average value of the similarity is calculated by the simple averaging circuit. This allows to obtain the average similarity score used in various filtering operation, objects detection tasks \cite{access}.

One of the possible applications of the proposed memristive detector is information encoding and cryptography. The dynamic system can be useful for encoding of the information in time, as set of video frames, which will be difficult to change due to the variability introduced by time domain switching. The decoder circuit for the inverse operation for this application should be also designed on hardware with sequencing circuits built using memristor logic gates \cite{logic}. The hardware implementation of such system would make an encryption process secure, as memristive threshold logic gates will be impossible to identify on chip, as the memristors are programmed to different resistance levels. The application of the circuit can also be extended to detect the other variations from the real time sensors, like the temperature variations.


The extension of this work will include the design of the delay circuit based on flux-charge capacitance devices and analog resistive memories for studying time dynamics of multiple fast moving objects.  In addition, the possibility of integration of the proposed system directly to the log-linear CMOS pixel arrays will be investigated. The memristor related problems will be analyzed, such as switching time, switching probability and different non-linear effects.

\section{Conclusion}
In this paper, we proposed a near pixel analog signal processing circuit for high speed object detection task in real-time settings. The system can be useful for static and dynamic application and can be integrated directly into image sensor. The functionality of the system is controlled by programming the memristor levels. The future work includes the system integration of various modules such as the delay circuit, pixel sensor and analog memory unit along with the processing circuits.


%


\bibliographystyle{IEEEtran}
\bibliography{ref}

\end{document}